# Theory of optical trapping of anti-reflection coating-coated spheres

Neng Wang[1,2,] Xiao Li[1], Jun Chen[3], Zhifang Lin[4], and Jack Ng[1,5*]

[1] *Department of Physics, Hong Kong Baptist University, Hong Kong, China*
[2] *Department of Physics, The Hong Kong University of Science and Technology, Hong Kong, China*
[3] *Institute of Theoretical Physics and Collaborative Innovation Center of Extreme Optics, Shanxi University, Shanxi, China*
[4] *Department of Physics, Fudan University, Shanghai, China*
[5] *Institute of Computational and Theoretical Studies, Hong Kong Baptist University, Hong Kong, China*

*Corresponding author: jacktfng@hkbu.edu.hk*

It was theoretically proposed and experimentally demonstrated that anti-reflection coating allows one to trap a high dielectric sphere, at the same time enhancing the transverse optical force. Here, by explicitly calculating the gradient force and the scattering force, we rigorously show that these were mainly consequences of the reduction in scattering force due to the suppression of backward scattering, and enhancement in gradient force due to the increased in overall particle size after coating. The reduction of scattering force can be understood within a ray optics theory and also the Mie theory. The coating approach only works for a spherical particle trapped by an aplanatic beam, and not in general.

## I. Introduction

Optical tweezers, the use of a tightly focused laser beam to trap small particles, have found fruitful applications across various scientific fields [1-4]. Particles in a laser beam subject to both scattering force $\mathbf{F}_s(\mathbf{x})$ and gradient force $\mathbf{F}_g(\mathbf{x})$ [5-10]. The former is divergence-less ($\nabla \cdot \mathbf{F}_s(\mathbf{x}) = 0$) and tends to axially push the particle away from the focus, while the latter is curl-less ($\nabla \times \mathbf{F}_g(\mathbf{x}) = \mathbf{0}$) and tends to attract the particle towards the focus. The particle can be trapped if the gradient force dominates [2]. Thus optical trapping can be improved by either reducing the optical scattering force or enhancing the optical gradient force. So far, these principles have been realized by either shaping the beam [11-14] or customizing the morphology of the trapped object [15, 16]. A simple and elegant example is to coat an anti-reflection coating (ARC) on a high dielectric spherical particle [17-19], as shown in Fig. 1. While the uncoated high dielectric particle is not trappable due to its high reflectivity, ARC could have made it trappable, which also leads to an unsurpassed transverse trap stiffness [18]. Nevertheless, despite the vigorous achievement [17-19], the underlining physics is still not fully understood. To our knowledge, there is no rigorous study on the scattering and gradient forces for a coated particle. This is partly due to the difficulty on decomposing the optical force into the scattering and gradient forces in the calculation and measurement.

In this letter, using the approach adopted by [10], we calculated the optical scattering and gradient forces for the uncoated and coated sphere illuminated by a Gaussian beam. Optical trapping is improved as the scattering force is reduced by the ARC as backward scattering is suppressed, while the gradient force is enhanced by ARC due to the increase in particle size.

**II. The scattering and gradient forces acting on uncoated and ARC coated spheres**

Consider a coated sphere, as shown in Fig. 1. It possesses inner radius $r_1$ and outer radius $r_2$. Its core, coating, and surrounding have refractive indices of $n_1$, $n_2$, and $n_0$, respectively. To make an ARC for a planar surface [20], the refractive indices must fulfill $\sqrt{n_1 n_0} \approx n_2$, and the radii must fulfill $n_2(r_2 - r_1) \approx 1/4\lambda$, where $\lambda$ is the vacuum wavelength. Throughout this letter, we consider the same materials as [18], namely, $n_0 = 1.33$ (water), $n_1 = 2.3$ (anatase titania), and $n_2 = 1.78$ (amorphous titania), which approximately fulfilled the ARC conditions. The trapping beam is an *x*-polarized and *z*-propagating fundamental Gaussian beam with vacuum wavelength $\lambda = 1.064$ μm and is focused by a high numerical aperture objective lens with N.A.=1.3. As shown in Fig. 1, the origin of the coordinate system is chosen to be the beam focus.

The total optical force $\mathbf{F}(\mathbf{x})$ acting on the sphere can be calculated rigorously using the Generalized-Lorenz-Mie theory and the Maxwell stress tensor [21-24]. Such an approach makes no approximation within classical electrodynamics and subjects only to numerical truncation error. The total force is numerically decomposed into the scattering force $\mathbf{F}_s(\mathbf{x})$ and gradient force $\mathbf{F}_g(\mathbf{x})$ using the approach presented in [10]:

$$\mathbf{F}_s(\mathbf{x}) = \int \frac{[\mathbf{q} \times \tilde{\mathbf{F}}(\mathbf{q})] \times \mathbf{q}/q^2}{(2\pi)^{3/2}} e^{i\mathbf{q}\cdot\mathbf{x}} d^3\mathbf{q},$$
$$\mathbf{F}_g(\mathbf{x}) = \int \frac{[\mathbf{q} \cdot \tilde{\mathbf{F}}(\mathbf{q})] \cdot \mathbf{q}/q^2}{(2\pi)^{3/2}} e^{i\mathbf{q}\cdot\mathbf{x}} d^3\mathbf{q},$$
(1)

where $\tilde{\mathbf{F}}(\mathbf{q})$ is the Fourier transform of the total optical force which is given by

$$\tilde{\mathbf{F}}(\mathbf{q}) = \int \frac{\mathbf{F}(\mathbf{x})}{(2\pi)^{3/2}} e^{-i\mathbf{q}\cdot\mathbf{x}} d^3\mathbf{x}.$$

In Fig. 2, we plotted the longitudinal (z direction) optical forces acting on the uncoated (solid lines) and ARC-coated (dotted lines) spheres when they are on the beam axis. There is no equilibrium for uncoated spheres, but equilibrium positions are created after introducing the ARC, in agreement with Refs 17-19. We plotted on Fig. 3 the gradient and scattering forces along the x and z directions for particles with and without the ARC ($r_1 = 0.5\mu m$) when they are on the xz plane. Comparing Figs. 3(a) and (b) with Figs. 3(e) and (f), we can see that with ARC, the scattering forces are reduced remarkably, especially when the particles are located around the beam focus, as shown in the blue spot at the center of Fig. 3(f). Comparing Fig. 3(c) and (d) with Figs. 3(g) and (h), the gradient forces are also enhanced. This is expected because the particle size increases after coating. For a large sphere ($r_1 = 5.0$ μm), we calculated the scattering and gradient forces and showed the results in Fig. 4. We can see that though the scattering force is greatly reduced, the gradient force is not enhanced much after the sphere is ARC coated, see more clearly in Fig. 5. Therefore, we remark for large particle, the weighing of the ARC in volume will be reduced, and thus effect of enhancing the gradient force should diminish. In short, the ARC-coated spheres are trapped because of the reduction in scattering force and enhancement in gradient force.

Note that the scattering force approaches zero at the beam focus for the coated particle shown in Fig. 1(f), accordingly the ARC-coated sphere could be trapped very close to the beam center where the transverse gradient force reaches its maximum. This explains why a large trapping force is realized [18]: first, the gradient force is enhanced mostly due to the increased particle size, second the reduced scattering force allows the particle to be trapped closer to the focus where the transverse gradient force is largest.

## III. The working mechanism of ARC

The function of the ARC can be understood approximately using the ray optics theory, which can be reasonably applied when the particle size is large compared with the wavelength, as the 5 micron sphere considered in this paper. As shown in Fig. 1(a), the beam is decomposed into a bunch of rays [25], for the particle located near the focus, all rays are normally incident onto the sphere surface. If the particle is sufficiently large such that the curvature of the sphere can be ignored, the reflectance for the rays will vanish as in the planar case. Due to the suppression of the back scattering, the forward part of the scattering force is significantly reduced. However, when the particle is located away from the beam focus, as shown in Fig. 1(b), the rays and the surface normal will not be parallel, and the angle between them varies with position. Consequently, no ARC could fit the position dependent angle. Accordingly, the back scattering will not be suppressed, and the mechanism for reducing the forward scattering force is no longer working. One can then infer that the ARC only works for an aplanatic lens, and it will not work for other type of incident field such as a plane wave, see detail in Appendix A.

The ray optics theory is simple and elegant, but it is inaccurate for particles with small size. Hereafter, we will apply a rigorous wave scattering theory. We introduce the normalized scattering intensity to describe the scattering feature [27, 28]:

$$S(\theta,\phi) = \lim_{kr \to \infty}(kr)^2 \frac{|\mathbf{E}_s(r,\theta,\phi)|^2}{|E_0|^2}, \qquad (2)$$

where $\mathbf{r} = (r, \theta, \phi)$ is spherical coordinate with origin coincided with the particle center, $k$ is the wavenumber in the medium, and $\mathbf{E}_s(r,\theta,\phi)$ denotes the scattered electric field. We shall only focus on the backward scattering ($\theta = \pi$). When the particle is located on the beam axis, Eq. (2) has a succinct analytic expression (see detail in Appendix B):

$$S(\pi,\phi) = |\sum_{n=1}^{\infty}(n+1)g_n(a_n - b_n)(\cos\phi\mathbf{e}_\theta - \sin\phi\mathbf{e}_\phi)|^2, \qquad (3)$$

where $a_n$ and $b_n$ are the Mie coefficients of the coated sphere,

$$g_n = \frac{\sqrt{2n+1}}{2i^n}\left[\frac{(n+1)i}{2n+1}j_{n-1}(ikz_c) + j_n(ikz_c) - \frac{ni}{2n+1}j_{n+1}(ikz_c)\right]\frac{kl_0}{e^{kl_0}}, \qquad (4)$$

is the beam shape coefficients [29], $l_0 = 1/2kw_0^2$ is the Rayleigh diffraction length, $w_0$ being the waist radius, and $z_c = l_0 - iz$ with $z$ being the location of sphere.

The Mie coefficients of the coated sphere are given by [27, 28]

$$a_n = \frac{\psi_n(y)[\psi_n'(\tilde{n}_2 y) - A_n \chi_n'(\tilde{n}_2 y)] - \tilde{n}_2 \psi_n'(y)[\psi_n(\tilde{n}_2 y) - A_n \chi_n(\tilde{n}_2 y)]}{\xi_n(y)[\psi_n'(\tilde{n}_2 y) - A_n \chi_n'(\tilde{n}_2 y)] - \tilde{n}_2 \xi_n'(y)[\psi_n(\tilde{n}_2 y) - A_n \chi_n(\tilde{n}_2 y)]},$$

$$b_n = \frac{\tilde{n}_2 \psi_n(y)[\psi_n'(\tilde{n}_2 y) - B_n \chi_n'(\tilde{n}_2 y)] - \psi_n'(y)[\psi_n(\tilde{n}_2 y) - B_n \chi_n(\tilde{n}_2 y)]}{\tilde{n}_2 \xi_n(y)[\psi_n'(\tilde{n}_2 y) - B_n \chi_n'(\tilde{n}_2 y)] - \xi_n'(y)[\psi_n(\tilde{n}_2 y) - B_n \chi_n(\tilde{n}_2 y)]}, \quad (5)$$

$$A_n = \frac{\tilde{n}_2 \psi_n(\tilde{n}_2 x)\psi_n'(\tilde{n}_1 x) - \tilde{n}_1 \psi_n'(\tilde{n}_2 x)\psi_n(\tilde{n}_1 x)}{\tilde{n}_2 \chi_n(\tilde{n}_2 x)\psi_n'(\tilde{n}_1 x) - \tilde{n}_1 \chi_n'(\tilde{n}_2 x)\psi_n(\tilde{n}_1 x)},$$

$$B_n = \frac{\tilde{n}_2 \psi_n(\tilde{n}_1 x)\psi_n'(\tilde{n}_2 x) - \tilde{n}_1 \psi_n'(\tilde{n}_1 x)\psi_n(\tilde{n}_2 x)}{\tilde{n}_2 \chi_n'(\tilde{n}_2 x)\psi_n(\tilde{n}_1 x) - \tilde{n}_1 \chi_n(\tilde{n}_2 x)\psi_n'(\tilde{n}_1 x)},$$

where $\psi_n(x) = x j_n(x)$, $\chi_n(x) = -x y_n(x)$, $\xi_n(x) = x h_n^{(1)}(x)$ are Ricatti-Bessel functions, $x = k r_1$ and $y = k r_2$ are dimensionless size parameters for the core and shell, respectively, and $\tilde{n}_1 = n_1 / n_0$ and $\tilde{n}_2 = n_2 / n_0$ are normalized refractive indices. For the ARC-coated sphere, we have $\tilde{n}_1 = \tilde{n}_2^2$ and $2\tilde{n}_2 (y - x) = \pi$. We first consider the large particle limit where $x \to \infty$, then the asymptotical formula for the Ricatti Bessel functions are introduced,

$$\xi_n(x) = \psi_n(x) - i\chi_n(x) \sim (-i)^{n+1} e^{ix},$$
$$\xi_n'(x) = \psi_n'(x) - i\chi_n'(x) \sim (-i)^n e^{ix}. \quad (6)$$

Using these expressions, the Mie coefficients reduce to a very simple expression and satisfy

$$a_n = b_n = \frac{\sin y + \cot(\tilde{n}_1 x) \cos y}{\cot(\tilde{n}_1 x) - i} e^{-iy}. \quad (7)$$

Eq. (7) indicates that for a very large ARC-coated sphere, Mie coefficients of all orders are the same. According to Eq. (3), the backward scattering goes to zero. For sphere with finite size, though Eq. (7) is not exact, low order Mie coefficients can be still approximated by Eq. (7) and fulfill $a_n \approx b_n$, see the numerical examples in Appendix C. While for the higher order terms, we note that for the Gaussian beam, $(n+1)g_n$ decreases rapidly with $n$, so the contributions from the high order terms in Eq. (3) are not important. Since the low order Mie coefficients satisfy $a_n \approx b_n$, the total backward scattering intensity almost vanishes, see Eq. (3). We remark that one could have reached a similar conclusion by noting that an impedance matched homogeneous spheres (with its permittivity equals to its permeability) will have $a_n = b_n$, where its reflection is expected to be low.

In Fig. 6 (a)-(c), we plotted the normalized scattering intensity for coated (blue dashed lines) and uncoated (red solid lines) spheres with different sizes when they are located at the Gaussian beam center. The uncoated high dielectric spheres have non-vanishing backward scatterings, which lead to a strong forward scattering force. But once the spheres are coated with the ARCs, though the total scattering efficiencies can be either enhanced or reduced (see the amplitude of the scattering intensity), the backward scatterings are always suppressed. However, when incident beam is a plane wave instead of a focusing beam, the ARC does not suppress the backward scattering, as shown in Fig. 6(d). This is because $(n+1)g_n = (n+1)\sqrt{n+1}/2$ for the plane wave, diverges with $n$. Accordingly, the higher order terms will contribute to backscattering, as they do not fulfill the low reflection condition of $a_n \approx b_n$. So the ARC does not work for a plane wave. In fact, according to its working principle, it should only work for an aplanatic beam.

**IV. ARC works better for large spheres**

Finally, we plotted the axial forces for spheres illuminated by Gaussian beams with different NA. For 0.5 micron radius particle shown in Fig. 7(a), sufficiently large NA is required for trapping even with ARC. In contrast, for the 5 micron radius particle shown in Fig. 7(b), ARC can greatly eliminate the forward forces, and make the high dielectric particle trappable. This seems to oppose our common sense that smaller particles are easier to trap. Using Eq. (3), we can also explain why ARC works better for large spheres. For smaller NA, $(n+1)g_n$ decreases slower as $n$ grows so that more higher order Mie coefficients will contribute to the backward scattering. As fewer multipole orders fulfill $a_n \approx b_n$ for smaller spheres while more orders do for larger spheres, ARC help reduce the backward scattering for larger spheres more efficiently, see also Fig 6 (a) and (c).

V. Summary

In summary, we have calculated and compared the scattering and gradient force for spheres with or without ARC illuminated by a Gaussian beam. The ARC can help reducing the forward scattering force when the sphere is located around the beam focus, so that the high dielectric spheres can be trapped very close to the beam focus. This is especially useful when the particle size is large so that its curvature approaches that of the planar case. For small particles, the increase in size after coating the ARC also lead to enhancement in optical force. Consequently, in agreement with previous studies, the trap stiffness is significantly enhanced. First, the gradient force increases as a consequence of the increase in particle size after coating. Second, the reduction in the forward scattering force due to the suppression of the backward scattering allows the particle to be trapped closer to the focus where the transverse gradient force is stronger. The former is more effective for larger sphere, while the latter is more effective for smaller sphere. Using ray optics, we have explained how the ARC works for the sphere trapped by a focused beam. We have also studied ARC using the Mie theory of the ARC-coated sphere and found that the vanishing backward scattering is a consequence of the special property owned by the ARC-coated sphere. Last but not least, reduced scattering force and enhanced gradient force might also improve optical manipulation in whatever areas where a conservative optical trap is needed.

**Acknowledgements**

The support of Hong Kong RGC through Grants No. HKBU 209913 and AoE/P-02/12 are gratefully acknowledged. The support from FRG2/16-17/095 is gratefully acknowledged.

**Appendix A. Optical scattering force for coated spheres in a Gaussian beam and plane wave**

Consider a coated sphere with inner and outer radii $r_1$ and, $r_2$, respectively. The refractive index of the core, coating, and background are $n_1$=2.3, $n_2$=1.78, and $n_0$=1.33, respectively. The optical gradient force vanishes when the particle is located in a plane wave or at the focus of a Gaussian beam. The scattering forces as functions of the coating thickness are shown in Fig. 1.

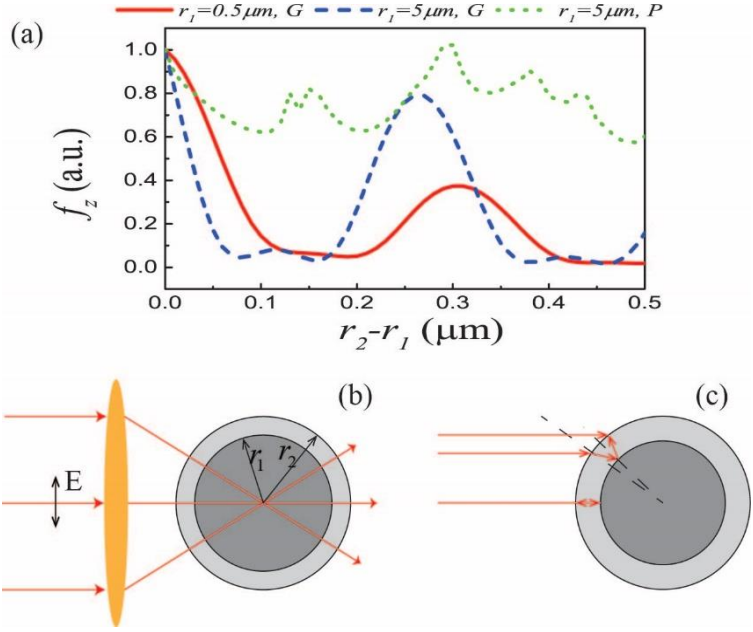

**Figure A1.** (a) The scattering force versus coating thickness for a particle located at the center of a linearly polarized fundamental Gaussian beam (red solid and blue dashed lines) and plane wave (green dotted line). The refractive indices of the core, coating and background medium are, respectively, $n_1 = 2.3, n_2 = 1.78, n_0 = 1.33$. The numerical aperture of the Gaussian beam is NA = 1.3, the wavelength is λ = 1.064 μm. (b) Schematic illustration of an ARC-coated sphere located at the focus of an aplanatic beam. (c) Schematic illustration of an ARC-coated sphere being illuminated by plane wave.

For the Gaussian beam, the scattering force is reduced significantly when the condition for antireflection coating (ARC) of a planar surface, $n_2(r_2 - r_1) \approx 1/4\lambda$ and $n_2 \simeq \sqrt{n_1 n_0}$, are fulfilled. However, for plane wave incidence, tuning the coating thickness does not reduce the scattering force, see green dotted line in Fig. A1(a).

As shown in Fig. A1(b), when the sphere is located at the beam center, each ray will normally incident on the sphere and the reflection is eliminated by the ARC. Therefore, back scattering is suppressed and so is the forward scattering force. But for the plane wave, the incident angles of the rays at different positions are different, see Fig. A1(b). Because of the incident angle-dependent phase differences, the reflected waves cannot be eliminated by ARC.

**Appendix B. Derivation of the backward scattering intensity (Eq. (3) in the main text)**

The normalized scattering intensity can describe the scattering feature of a particle illuminated

by a beam, which is given by [26, 27]

$$S(\theta,\phi) = \lim_{kr\to\infty}(kr)^2 \frac{|\mathbf{E}_s(r,\theta,\phi)|^2}{|E_0|^2}, \tag{B1}$$

where $\mathbf{r}=(r,\theta,\phi)$ is the spherical coordinate centered at the particle center, $k$ is the wavenumber in the medium, and $\mathbf{E}_s(r,\theta,\phi)$ denotes the scattered electric field. According to the Mie theory [26], the incident and scattered fields can be written as

$$\begin{aligned}\mathbf{E}_i(\mathbf{r}) &= \sum_{mn} i^{n+1} E_0 [p_{mn}\mathbf{N}_{mn}^{(1)}(k,\mathbf{r}) + q_{mn}\mathbf{M}_{mn}^{(1)}(k,\mathbf{r})],\\ \mathbf{E}_s(\mathbf{r}) &= \sum_{mn} i^{n+1} E_0 [a_{mn}\mathbf{N}_{mn}^{(3)}(k,\mathbf{r}) + b_{mn}\mathbf{M}_{mn}^{(3)}(k,\mathbf{r})],\end{aligned} \tag{B2}$$

where $p_{mn}$ and $q_{mn}$ are the beam shape coefficients [21-23], $a_{mn}$ and $b_{mn}$ are scattering coefficients, which are related by the Mie coefficients [26]: $a_{mn}=a_n p_{mn}$ and $b_{mn}=b_n q_{mn}$, and $\mathbf{N}_{mn}^{(J)}$ and $\mathbf{M}_{mn}^{(J)}$ are the vector spherical wave functions given by

$$\mathbf{N}_{mn}^{(J)}(k,\mathbf{r}) = [\tau_{mn}(\cos\theta)\mathbf{e}_\theta + i\pi_{mn}(\cos\theta)\mathbf{e}_\phi]\frac{\xi_n'(kr)}{kr}e^{im\phi} + \mathbf{e}_r n(n+1)C_{mn}P_n^m(\cos\theta)\frac{\xi_n(kr)}{(kr)^2}e^{im\phi},$$

$$\mathbf{M}_{mn}^{(J)}(k,\mathbf{r}) = [i\pi_{mn}(\cos\theta)\mathbf{e}_\theta - \tau_{mn}(\cos\theta)\mathbf{e}_\phi]\frac{\xi_n(kr)}{kr}e^{im\phi},$$

where $C_{mn}=\sqrt{\frac{(2n+1)}{n(n+1)}\frac{(n-m)!}{(n+m)!}}$, $\xi_n(kr)$ is the Ricatti Bessel functions, which is $\xi_n(kr)=(kr)j_n(kr)$ when $J=1$ and $\xi_n(kr)=(kr)h_n^{(1)}(kr)$ when $J=3$, and $P_n^m(\cos\theta)$ is the associated Legendre function of the first kind:

$$P_n^m(x) = \frac{1}{2^n n!}(1-x^2)^{m/2}\frac{d^{n+m}}{dx^{n+m}}[(x^2-1)^n], \quad P_n^{-m}(x)=(-1)^m\frac{(n-m)!}{(n+m)!}P_n^m(x). \tag{B3}$$

The two auxiliary angular functions $\pi_{mn}(\cos\theta)$ and $\tau_{mn}(\cos\theta)$ are defined as

$$\pi_{mn}(\cos\theta) = C_{mn}\frac{m}{\sin\theta}P_n^m(\cos\theta), \quad \tau_{mn}(\cos\theta)=C_{mn}\frac{d}{d\theta}P_n^m(\cos\theta),$$

which satisfy [29]

$$\pi_{-mn}(x)=(-1)^{m+1}\pi_{mn}(x), \quad \tau_{-mn}(x)=(-1)^m\tau_{mn}(x). \tag{B4}$$

For linearly polarized Gaussian beam, the beam shape coefficients haves analytic expressions when the particle is located on the beam axis [28]:

$$p_{1n} = g_n = \frac{\sqrt{2n+1}}{2i^n} \left[ \frac{(n+1)i}{2n+1} j_{n-1}(ikz_c) + j_n(ikz_c) - \frac{ni}{2n+1} j_{n+1}(ikz_c) \right] \frac{kl_0}{e^{kl_0}},$$

$$p_{1n} = -p_{-1n} = q_{1n} = q_{-1n},$$

$$p_{mn} = q_{mn} = 0, \quad m \neq \pm 1.$$

(B5)

where $l_0 = 1/2kw_0^2$ is the Rayleigh diffraction length with $w_0$ being the waist radius, and $z = l_0 - iz_0$ with $z_0$ being the location of beam center. As one can see from Eq. (B5), for both $p_{mn}$ and $q_{mn}$, only the azimuthal modes with $m = \pm 1$ contribute. Then the scattered field in the far field ($kr \to \infty$) is given by

$$\begin{aligned}\lim_{kr\to\infty}(kr)\mathbf{E}_s &= \lim_{kr\to\infty}(kr)\sum_n i^{n+1} E_0 \left[ a_n (p_{-1n}\mathbf{N}^{(3)}_{-1n} + p_{1n}\mathbf{N}^{(3)}_{1n}) + b_n (q_{-1n}\mathbf{M}^{(3)}_{-1n} + q_{1n}\mathbf{M}^{(3)}_{1n}) \right] \\
&= \lim_{kr\to\infty}(kr)\sum_n i^{n+1} E_0 g_n \left[ a_n(-\mathbf{N}^{(3)}_{-1n} + \mathbf{N}^{(3)}_{1n}) + b_n(\mathbf{M}^{(3)}_{-1n} + \mathbf{M}^{(3)}_{1n}) \right] \\
&= \lim_{kr\to\infty} 2\sum_n i^{n+1} E_0 g_n \begin{bmatrix} a_n([\tau_{1n}(x)\cos\phi\mathbf{e}_\theta - \pi_{1n}(x)\sin\phi\mathbf{e}_\phi]\xi_n'(kr) \\ +ib_n(\pi_{1n}(x)\cos\phi\mathbf{e}_\theta - \tau_{1n}(x)\sin\phi\mathbf{e}_\phi)\xi_n(kr) \end{bmatrix}.\end{aligned}$$

(B6)

Using the asymptotical formula for the Ricatti Bessel functions for $kr \to \infty$,

$$\xi_n(kr) \sim (-i)^{n+1} e^{ikr}, \quad \xi_n'(kr) \sim (-i)^n e^{ikr} = i\xi_n(kr),$$

Eq. (B6) reduces to

$$\lim_{kr\to\infty}(kr)\mathbf{E}_s = \lim_{kr\to\infty} 2\sum_n iE_0 g_n \begin{bmatrix} a_n([\tau_{1n}(x)\cos\phi\mathbf{e}_\theta - \pi_{1n}(x)\sin\phi\mathbf{e}_\phi] \\ +b_n(\pi_{1n}(x)\cos\phi\mathbf{e}_\theta - \tau_{1n}(x)\sin\phi\mathbf{e}_\phi) \end{bmatrix} e^{ikr}.$$

(B7)

For the backward direction, namely $x = \cos\pi = -1$, the auxiliary functions read

$$\pi_{1n} = -\tau_{1n} = (-1)^{n+1}\frac{\sqrt{n+1}}{2}.$$

(B8)

Then substituting Eq. (B7) and Eq. (B8) into Eq. (B1), we have the backward scattering intensity as

$$S(\theta,\phi) = |\sum_n (n+1)g_n(a_n - b_n)(\cos\phi\mathbf{e}_\theta - \sin\phi\mathbf{e}_\phi)|^2,$$

(B9)

which is Eq. (3) in the main text.

**Appendix C. Numerical examples of Mie coefficients of the antireflection-coated spheres**

The Mie coefficients for a coated sphere are given by [26, 27]

$$a_n = \frac{\psi_n(y)[\psi_n'(n_2 y) - A_n \chi_n'(n_2 y)] - n_2 \psi_n'(y)[\psi_n(n_2 y) - A_n \chi_n(n_2 y)]}{\xi_n(y)[\psi_n'(n_2 y) - A_n \chi_n'(n_2 y)] - n_2 \xi_n'(y)[\psi_n(n_2 y) - A_n \chi_n(n_2 y)]},$$

$$b_n = \frac{n_2 \psi_n(y)[\psi_n'(n_2 y) - A_n \chi_n'(n_2 y)] - \psi_n'(y)[\psi_n(n_2 y) - A_n \chi_n(n_2 y)]}{n_2 \xi_n(y)[\psi_n'(n_2 y) - A_n \chi_n'(n_2 y)] - \xi_n'(y)[\psi_n(n_2 y) - A_n \chi_n(n_2 y)]},$$

$$A_n = \frac{n_2 \psi_n(n_2 x) \psi_n'(n_1 x) - n_1 \psi_n'(n_2 x) \psi_n(n_1 x)}{n_2 \chi_n(n_2 x) \psi_n'(n_1 x) - n_1 \chi_n'(n_2 x) \psi_n(n_1 x)},$$

$$B_n = \frac{n_2 \psi_n(n_1 x) \psi_n'(n_2 x) - n_1 \psi_n'(n_1 x) \psi_n(n_2 x)}{n_2 \chi_n'(n_2 x) \psi_n(n_1 x) - n_1 \chi_n(n_2 x) \psi_n'(n_1 x)},$$

(C1)

where $\psi_n(x) = x j_n(x), \chi_n(x) = -x y_n(x), \xi_n(x) = x h_n^{(1)}(x)$ are Ricatti-Bessel functions, $x = k r_1, y = k r_2$ are dimensionless size parameters for the core and shell, respectively, and $\tilde{n}_1 = n_1 / n_0, \tilde{n}_2 = n_2 / n_0$ are normalized refractive indices. When the shell is an ARC, the low order Mie coefficients for large sphere can be approximated as

$$a_n = b_n = \frac{\sin y + \cot(\tilde{n}_1 x) \cos y}{\cot(\tilde{n}_1 x) - i} e^{-iy}. \tag{C2}$$

Tables 1 and 2 show the Mie coefficients calculated by Eq. (C1) for ARC-coated spheres of different sizes, and $(n+1)g_n$ calculated by Eq. (B5). The particle is located at the focus. For the smaller one micron sphere, though the Mie coefficients cannot be approximated by Eq. (C2) very well, low order Mie coefficients still satisfy $a_n \approx b_n$. For the larger 5 micron sphere, low order Mie coefficients are well approximated by Eq. (C2) and satisfy $a_n \approx b_n$. Moreover, $(n+1)g_n$ diminishes quickly with increasing $n$. Consequently, the backward scattering is dominated by the low order terms which has a vanishing contribution to the backward scattering.

| $n$ | $a_n$ | $b_n$ | $(n+1)g_n$ |
|---|---|---|---|
| 1 | 0.036436+0.187373i | 0.0487883+0.215425 i | 1.32691 |
| 2 | 0.101541+0.302044i | 0.0802716+0.271713 i | 1.53799 |
| 3 | 0.186108+0.389194i | 0.223248+0.416423 i | 1.16625 |
| 4 | 0.417332+0.493119i | 0.360582+0.480169i | 0.655511 |
| 5 | 0.656048+0.475025i | 0.69021+0.462407i | 0.291314 |
| 6 | 0.93799+0.241173i | 0.962933+0.188926i | 0.106639 |
| 7 | 0.961466-0.192481i | 0.999996+0.00189196i | 0.0330921 |

**Table 1**. Mie coefficients for an ARC-coated sphere with inner and outer radii $r_1 = 1.0$ μm and

$r_2 = 1.15$ μm, respectively. The sphere is located at the focus of a linearly polarized Gaussian beam. Mie coefficients calculated by Eq. (C2) is 0.0272428+0.16279i.

| $n$ | $a_n$ | $b_n$ | $(n+1)g_n$ |
|---|---|---|---|
| 1 | 0.455965-0.498057i | 0.482446-0.499692i | 1.3392 - 0.0705567i |
| 2 | 0.461564-0.49852i | 0.43433-0.495669i | 1.56435 - 0.232702i |
| 3 | 0.402758-0.490453i | 0.431044-0.495222i | 1.17898 - 0.330319i |
| 4 | 0.389207-0.48757i | 0.360061-0.480018i | 0.642546 - 0.286775i |
| 5 | 0.310372-0.462646i | 0.340283-0.473804i | 0.267418 - 0.17560i |
| 6 | 0.279918-0.448959i | 0.250415-0.433252i | 0.0874659 - 0.0819216i |
| 7 | 0.190427-0.392638i | 0.218725-0.413382i | 0.0226966 - 0.0305472i |

**Table. 2**. Mie coefficients for an ARC-coated sphere with inner and outer radii $r_1 = 5.0$ μm and $r_2 = 5.15$ μm, respectively. The sphere is located at the focus of a linearly polarized Gaussian beam. Mie coefficients calculated by Eq. (C2) is 0.452864 - 0.497773i.

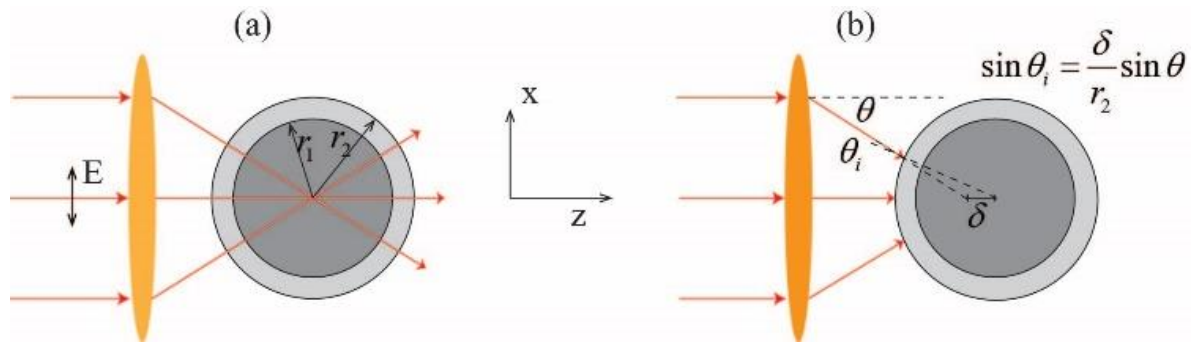

Fig. 1. Schematic illustration of an ARC-coated sphere located (a) at and (b) off the focus of an aplanatic beam.

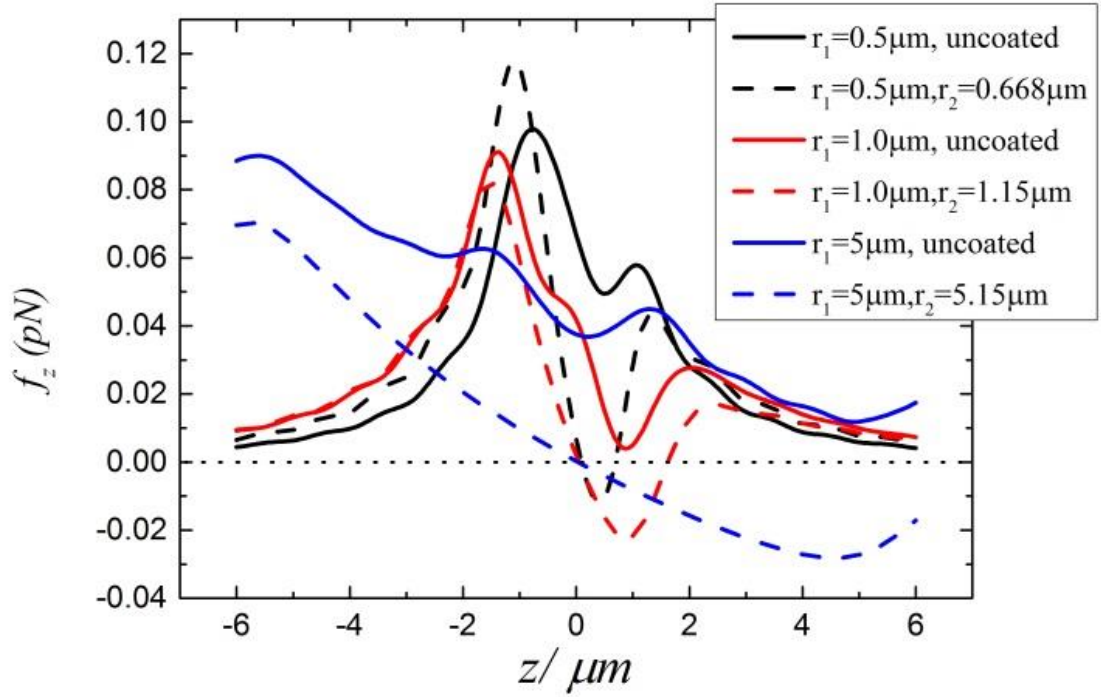

Fig. 2. The longitudinal optical forces acting on the uncoated (solid lines) and ARC-coated (dotted lines) spheres in a Gaussian beam. The spheres are located on the beam axis. and $E_0 = 10^6 V/m$.

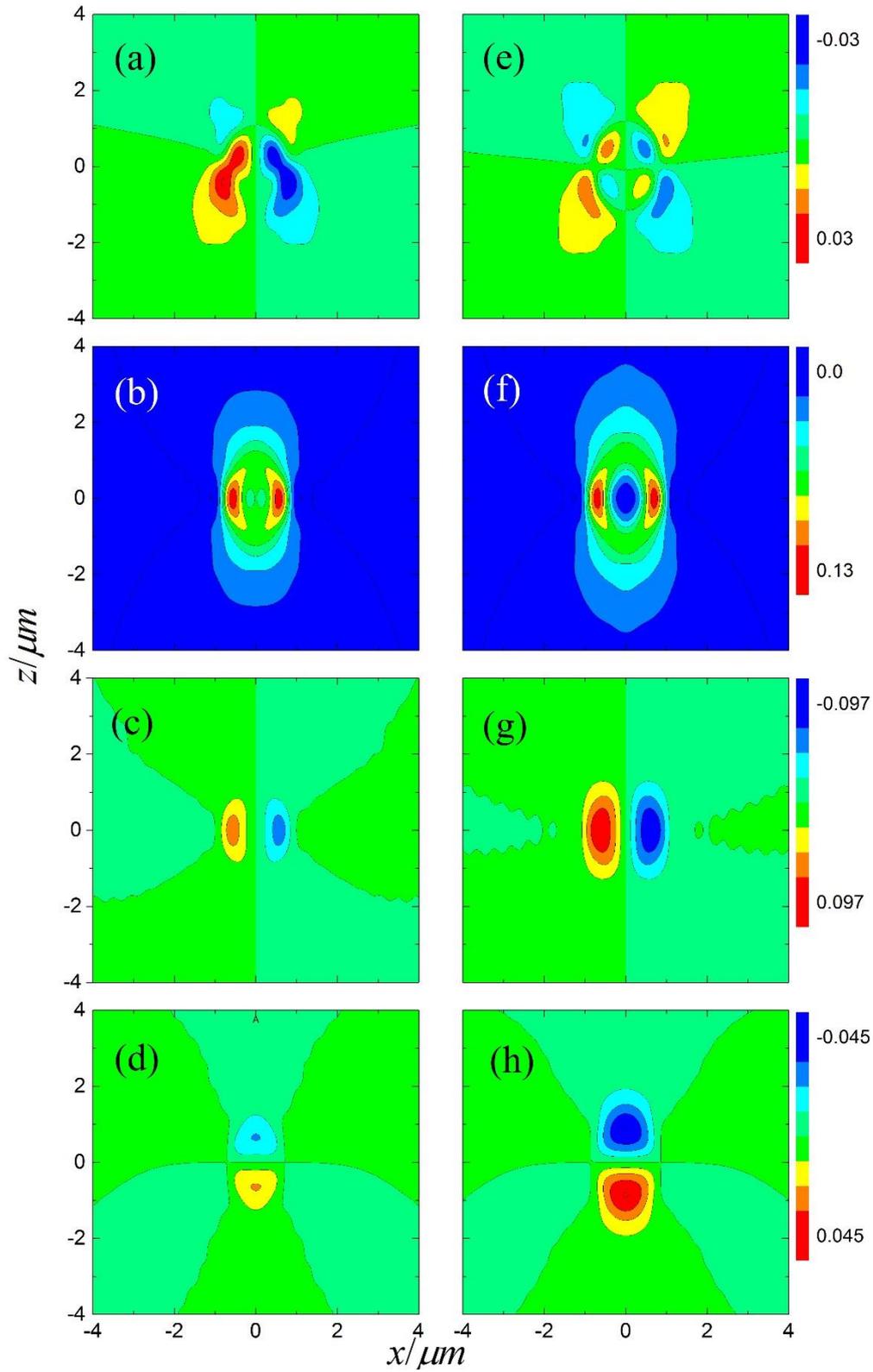

Fig. 3. Gradient and scattering force for uncoated (left column) and ARC-coated (right column) spheres illuminated by an *x*-polarized Gaussian beam. The first, second, third, and forth rows are, respectively,

$(\mathbf{F}_s)_x$, $(\mathbf{F}_s)_z$, $(\mathbf{F}_g)_x$, and $(\mathbf{F}_g)_z$. The inner and outer radii of the sphere are $r_1 = 0.5$ μm, $r_2 = 0.668$ μm.

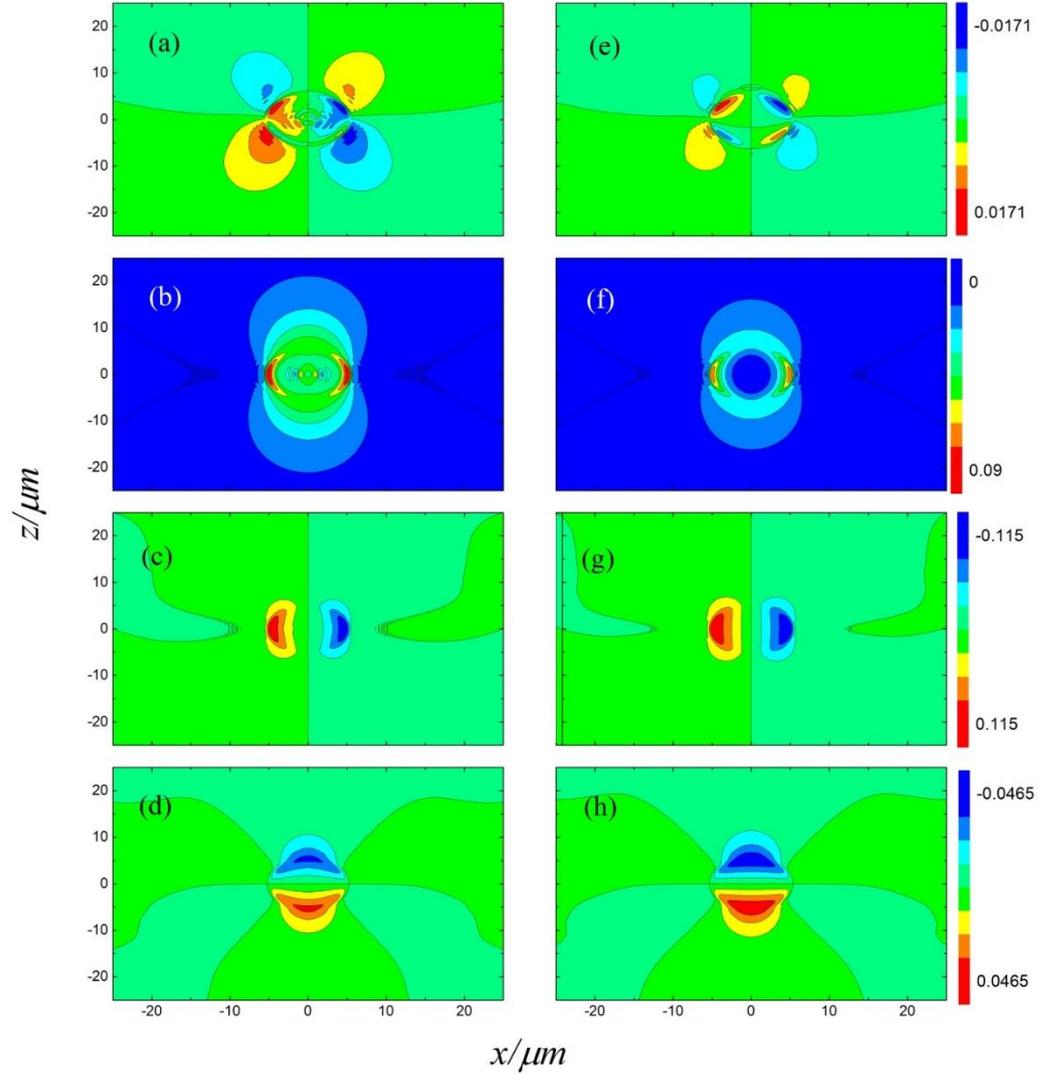

Fig. 4. Gradient and scattering force for uncoated (left column) and ARC-coated (right column) illuminated by an *x*-polarized Gaussian beam. The first, second, third, and forth rows are, respectively, $(\mathbf{F}_s)_x$, $(\mathbf{F}_s)_z$, $(\mathbf{F}_g)_x$, and $(\mathbf{F}_g)_z$. The unit of optical force is pN. The inner and outer radii of the sphere are $r_1 = 5.0$ μm and $r_2 = 5.15$ μm, respectively. Here, $E_0 = 10^6 V/m$.

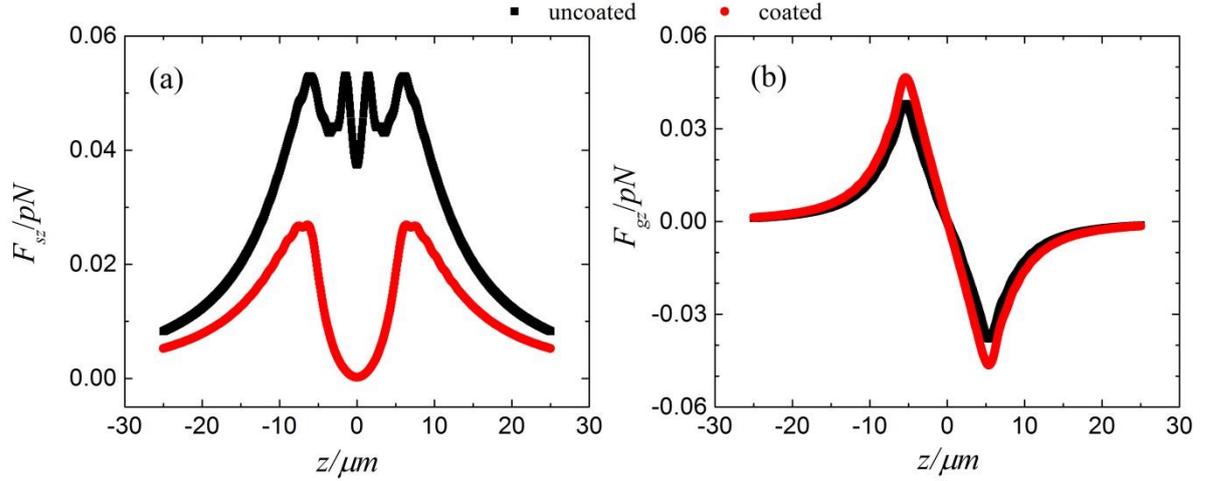

Fig. 5. The longitudinal component of scattering (a) and gradient (b) forces for uncoated (black) and ARC-coated (red) spheres. The parameters used are the same with Fig. 6.

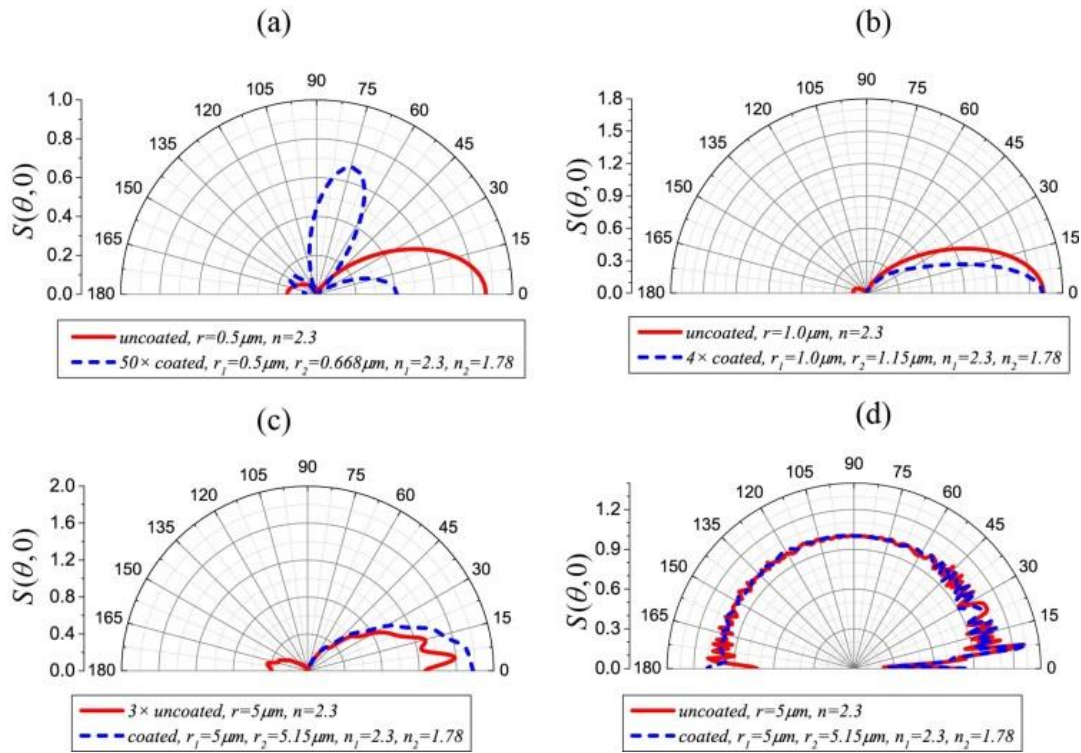

Fig. 6. Normalized scattering intensities for the uncoated (red solid lines) and antireflection-coated (blue dashed lines) spheres with different sizes in a Gaussian beam [(a)-(c)] and a plane wave [(d)].

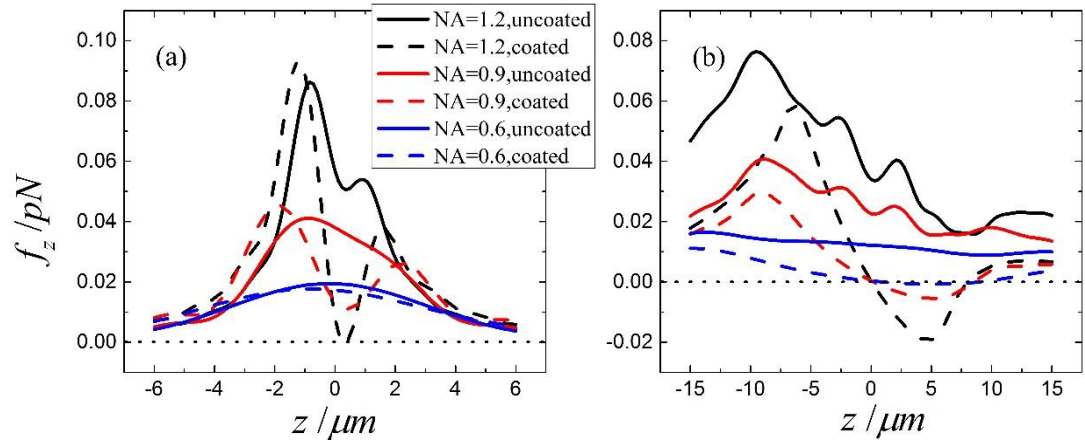

Fig. 7. Axial optical forces for beams with different NA. (a) is for core radius 0.5 μm, and (b) is for core radius 5.0 μm.